\documentclass[preprint,aps,nofootinbib]{revtex4}

\usepackage{graphicx}
\usepackage{dcolumn}
\usepackage{bm}
\usepackage{epsfig}
\usepackage{graphicx}
\usepackage[bookmarks=true]{hyperref}
\usepackage[caption=false]{subfig}

\usepackage{amsmath}
\usepackage{amsfonts}
\usepackage{amssymb}
\usepackage[section]{placeins}  

\usepackage{amsmath}
\usepackage{amsfonts}
\usepackage{amssymb}
\usepackage{color}

\def\be{\begin{equation}}
\def\ee{\end{equation}}
\def\bea{\begin{eqnarray}}
\def\eea{\end{eqnarray}}

\def\ss2l{SS2$\ell$}
\def\3l{3$\ell$}

\def\slashchar#1{\setbox0=\hbox{$#1$}           
   \dimen0=\wd0                                 
   \setbox1=\hbox{/} \dimen1=\wd1               
   \ifdim\dimen0>\dimen1                        
      \rlap{\hbox to \dimen0{\hfil/\hfil}}      
      #1                                        
   \else                                        
      \rlap{\hbox to \dimen1{\hfil$#1$\hfil}}   
      /                                         
   \fi}

\begin{document}

\title{A Double Take on New Physics  in Double Higgs Production}
\vspace*{1cm}

\author{\vspace{1cm} Chuan-Ren Chen$^{\, a}$ and   Ian Low$^{\, b,c}$ }

\affiliation{
\vspace*{.5cm}
  \mbox{$^a$Department of Physics, National Taiwan Normal University, Taipei 116, Taiwan}\\
\mbox{$^b$ Department of Physics and Astronomy, Northwestern University, Evanston, IL 60208, USA} \\
\mbox{$^c$ High Energy Physics Division, Argonne National Laboratory, Argonne, IL 60439, USA}\\
\vspace*{1cm}}

\begin{abstract}
\vspace*{0.5cm}

Gluon-initiated double Higgs production is the most important channel to extract the Higgs self-coupling at hadron colliders. However, new physics could enter into this channel in several distinctive ways including, but not limited to, the Higgs self-coupling, a modified top Yukawa coupling, and an anomalous Higgs-top quartic coupling. In this work we initiate a  study on the interplay of these effects in the kinematic distributions of the Higgs bosons. More specifically, we divide  the $p_T$ and the total invariant mass  spectra into two bins and use the differential rates in each bin to constrain the magnitude of the  aforementioned effects.  Significantly improved results could be obtained over those using total cross section alone. However, some degeneracy remains, especially in the determination of the Higgs trilinear coupling. Therefore, an accurate measurement of the Higgs self-coupling in this channel would require precise knowledge of the magnitudes of  other new physics effects. We base our analysis on a future $pp$ collider  at $\sqrt{s}=100$ TeV.

\end{abstract}


\maketitle

\section{Introduction}

\begin{figure}[t]
\includegraphics[scale=0.7, angle=0]{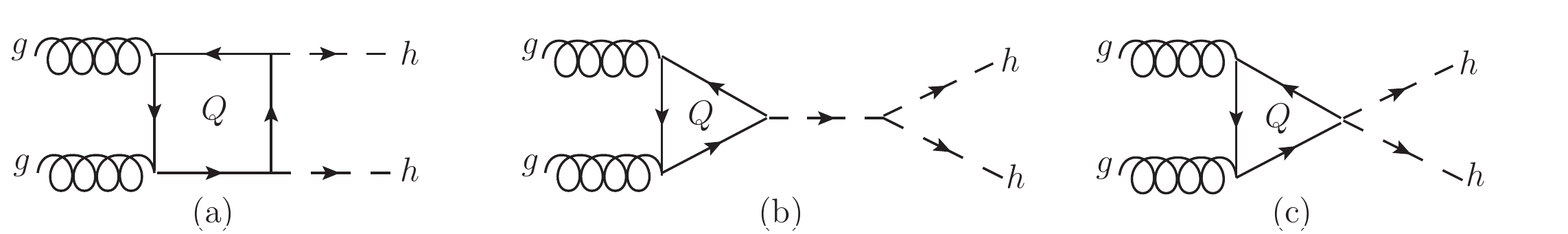}  
\caption{\label{fig1}{\em Feynman diagrams contributing to double Higgs production at hadron colliders.
}}
\end{figure}

Self-interaction is the only aspect of the newly discovered 125 GeV Higgs boson \cite{Aad:2012tfa,Chatrchyan:2012ufa} that has not been measured experimentally. Yet these interactions represent  the only window to reconstruct the scalar potential of the Higgs boson and directly test the underlying framework of spontaneous  symmetry breaking through a scalar vacuum expectation value (VEV). In hadron colliders gluon-initiated double Higgs production, $gg\to h h$ \cite{Glover:1987nx, Plehn:1996wb}, is typically employed to measure the Higgs self-coupling \cite{Baur:2002rb}. The standard model (SM) expectation for this production rate is only 33.9 fb at the 14 TeV Large Hadron Collider \cite{HXWG}, making such a measurement challenging unless the rate is strongly enhanced. Part of the reason for such a small rate is a strong cancellation near the kinematic threshold \cite{Li:2013rra} between the two contributing diagrams in the SM, which are the box diagram in Fig.~\ref{fig1}a and the triangle diagram in Fig.~\ref{fig1}b.  However,  at a 100 TeV $pp$ collider the SM rate increases dramatically to 1.42 pb \cite{HXWG} due to the growing  luminosity in the gluon parton distribution function (PDF) at smaller Bjorken $x$, thereby providing an opportunity to reconstruct the Higgs scalar potential with precision \cite{Baglio:2012np,Yao:2013ika}.

While it is of great importance to verify that the electroweak symmetry is indeed broken spontaneously by a scalar VEV, the ultimate goal of any such measurement is to discover new physics beyond the SM. It then becomes imperative to analyze the double Higgs production in a broad context, by considering various possible new physics that could enter into this particular channel. With this mindset, it was realized that significant effects could result from a new diagram, which is shown in Fig.~\ref{fig1}c,  involving the anomalous Higgs-top quartic coupling of the form $\bar{t}thh$ \cite{Dib:2005re,Grober:2010yv}. When allowing for the presence of such a coupling, it was found in Ref.~\cite{Contino:2012xk}   the total production cross section is the least sensitive to the Higgs self coupling, making a measurement of this coupling especially challenging.

Recently there has been much attention  on new physics in double Higgs productions \cite{Dib:2005re,Grober:2010yv, Contino:2012xk,Dawson:2012mk,Gillioz:2012se,Wang:2007zx}, however, the majority, if not all, focused only on using the total rate measurement. In the present work we initiate a study to disentangle different new physics effects in the double Higgs production using  kinematic distributions of the Higgs bosons. 
In particular, we focus on $m_{hh}$, the total invariant mass, and $p_T$ spectra of the Higgs and study the interplay of various new physics effects in these kinematic distributions. 

This work is organized as follows. In the next Section we introduce a parameterization of new physics effects in the differential spectra of double Higgs production. Then in Section III we study the impact of the new physics effects on the kinematic distributions, which is followed by a numerical study on constraints from using the kinematic information in a 100 TeV $pp$ collider. In Section IV we present the conclusions.

\section{New Physics in double Higgs production}
\label{sect:new}

SM contributions to double Higgs production have been calculated long ago in Refs.~\cite{Glover:1987nx,Plehn:1996wb}, while the additional contribution from the anomalous Higgs-top coupling was studied in Refs.~\cite{Dib:2005re,Grober:2010yv}. Using these results,  we write the partonic differential cross-section from the three diagrams in Fig.~\ref{fig1}  as
\bea
\label{eq:amp}
&&\frac{d\hat{\sigma}(gg\to hh)}{d\hat{t}} =  \frac{G_F^2 \alpha_s^2}{512(2\pi)^3}\nonumber \\
&& \qquad \times \left[ \left|\left(g_{3h} \frac{1}{\hat{s}-m_h^2}\ g_{htt} + g_{hhtt}  \right)  \frac{v^2}{m_t}  {F}_{\bigtriangleup}  + g_{htt}^2\frac{v^2}{m_t^2} {F}_\Box \right|^2 + \left| g_{htt}^2\frac{v^2}{m_t^2} {G}_\Box \right|^2 \right] \ ,
\eea
 where $g_{3h}$ is the trilinear Higgs coupling, $g_{htt}$ is the Higgs coupling to $t\bar{t}$,  and $g_{hhtt}$ is the anomalous Higgs-top coupling. These couplings appear in the Lagrangian as
\be
\label{eq:lagcoupling}
 \frac1{3!} g_{3h}\, h^3 + g_{htt}\, h\bar{t}t + \frac1{2!} g_{hhtt}\, h^2 \bar{t}t \ .
\ee
Therefore in the SM we have
\be
g_{3h}^{\rm (SM)} = \frac{3m_h^2}{v} \ , \qquad g_{htt}^{\rm (SM)} = \frac{m_t}{v}\ , \qquad g_{hhtt}^{\rm (SM)}= 0\ ,
\ee
where $v=246$ GeV is the Higgs vacuum expectation value. In the above, $F_\bigtriangleup$, $F_\Box$, and $G_\Box$ are loop functions depending on partonic Mandelstam  variables $\hat{s}$, $\hat{t}$, $\hat{u}$ and the mass of the fermion running in the loop.   Analytical expressions of them can be found in, for example, Ref.~\cite{Plehn:1996wb}, whose notations we follow. In addition, $\alpha_s$ is the strong coupling constant and $G_F=1/(\sqrt{2}v^2)$ is the Fermi constant.

The expression in Eq.~(\ref{eq:amp}) is quite general and captures effects from new physics in a wide class of models. In particular, if there exists new colored fermions with significant couplings to the Higgs, their contributions to $gg\to hh$ could be included by computing the Higgs couplings in the mass eigenbasis and using the  mass eigenvalues in the loop functions. In this work we only include the SM top quark in the loop functions and focus on the  interplay of effects from terms in Eq.~(\ref{eq:lagcoupling}), as effects from new colored fermions have been studied closely in Refs.~\cite{Dib:2005re, Dawson:2012mk}. In the SM Eq.~(\ref{eq:amp}) reduces to 
\bea
\label{eq:ampSM}
  \frac{G_F^2 \alpha_s^2}{512(2\pi)^3}\left[ \left|\frac{3m_h^2}{\hat{s}-m_h^2}  F_{\bigtriangleup}  +  F_\Box \right|^2 + \left| G_\Box \right|^2\right] \ .
\eea
Notice we have included a factor of 1/2 for identical particles in the final state that was missing in some literature. Our result agrees with that in Ref.~\cite{Dawson:2012mk}.

It is convenient to parameterize Eq.~(\ref{eq:amp}) with three dimensionless coefficients
\bea
\label{eq:ampara}
\frac{d\hat{\sigma}(gg\to hh)}{d\hat{t}} &=&  \frac{G_F^2 \alpha_s^2}{512(2\pi)^3}\left[ \left|\left(c_{tri} \frac{3m_h^2}{\hat{s}-m_h^2} + c_{nl} \right) F_{\bigtriangleup}  + c_{box} F_\Box \right|^2 + \left| c_{box} G_\Box \right|^2\right] \ ,
\eea
so that
\be
c_{box}^{\rm (SM)}= 1 \ , \qquad c_{tri}^{\rm (SM)} = 1 \ , \qquad c_{nl}^{\rm (SM)} = 0 \ .
\ee
The mapping between these coefficients and the relevant Higgs couplings is simple\footnote{In terms of the notations in Ref.~\cite{Contino:2012xk}, we have $c_{tri} = c\, d_3$, $c_{nl} = 2 c_2$, and $c_{box} = c^2$.}
 \be
 c_{tri} = g_{3h}\,g_{htt}\, \frac{v^2}{3m_h^2 m_t}\ , \qquad c_{nl} = g_{hhtt}\,\frac{v^2}{m_t} \ , \qquad c_{box} = \left(g_{htt}\,\frac{v}{m_t}\right)^2 \ .
 \ee
In the framework of effective theory, new physics enters into low-energy Higgs observables only through  gauge-invariant operators of dimension-6 or higher. Thus  we expect
\be
\label{eq:power}
\delta c_{tri, box, nl} \sim {\cal O}\left(\frac{v^2}{\Lambda_{\rm new}^2}\right) \ ,
\ee
where $\Lambda_{\rm new}$ represents the generic scale of new physics. In this work we will adopt a bottom-up approach by allowing all three coefficients to vary freely, without being constrained by the power counting in Eq.~(\ref{eq:power}).

Fig.~\ref{fig1}b and Fig.~\ref{fig1}c have the same loop function as in the single Higgs production from the gluon fusion. Throughout this study we only include the top quark in the heavy quark loop.  It is known that the $m_t\to \infty$ limit works  well in $F_\bigtriangleup$ and  terribly in $F_\Box$ and $G_\Box$ \cite{Gillioz:2012se, Dawson:2012mk}. As a result, the celebrated low-energy Higgs theorems \cite{Ellis:1975ap} cannot apply in the double Higgs production and it is important to keep the full $m_t$ dependence. Heuristically this is due to the fact that the partonic center-of-mass (CM) energy in the double Higgs production must always be above the kinematic threshold: $\hat{s} \ge 4m_h^2$, while the low-energy theorems require $\hat{s} \ll 4m_t^2$ \cite{Dawson:2012mk}. Therefore, scenarios with new colored particles must be treated with care, by including the full mass dependence in the loop functions.

\section{Kinematic Distributions}

In a hadron collider, the leading order (LO) differential cross-section in the laboratory frame can be obtained by convoluting the partonic cross-section with the gluon PDF's:
\be
\frac{d^2\sigma(pp\to hh)}{dm_{hh}\, dp_T} =\int_\tau^1 \frac{dx}{x}g(x,\mu_F) g\left(\frac{\tau}{x},\mu_F\right) \frac{2m_{hh}}{s} \frac{d\hat{\sigma}(gg\to hh)}{dp_T} \ ,
\ee
where $s$ is the hadronic CM energy, $m_{hh}=\sqrt{\hat{s}}$,  $\tau=\hat{s}/s$, and $p_T$ is the transverse momentum of the Higgs boson:
\be
p_T^2 = \frac{\hat{u}\hat{t} - m_h^4}{\hat{s}} \ .
\ee 

\begin{figure}[t]
\subfloat[]{\includegraphics[scale=0.67, angle=0]{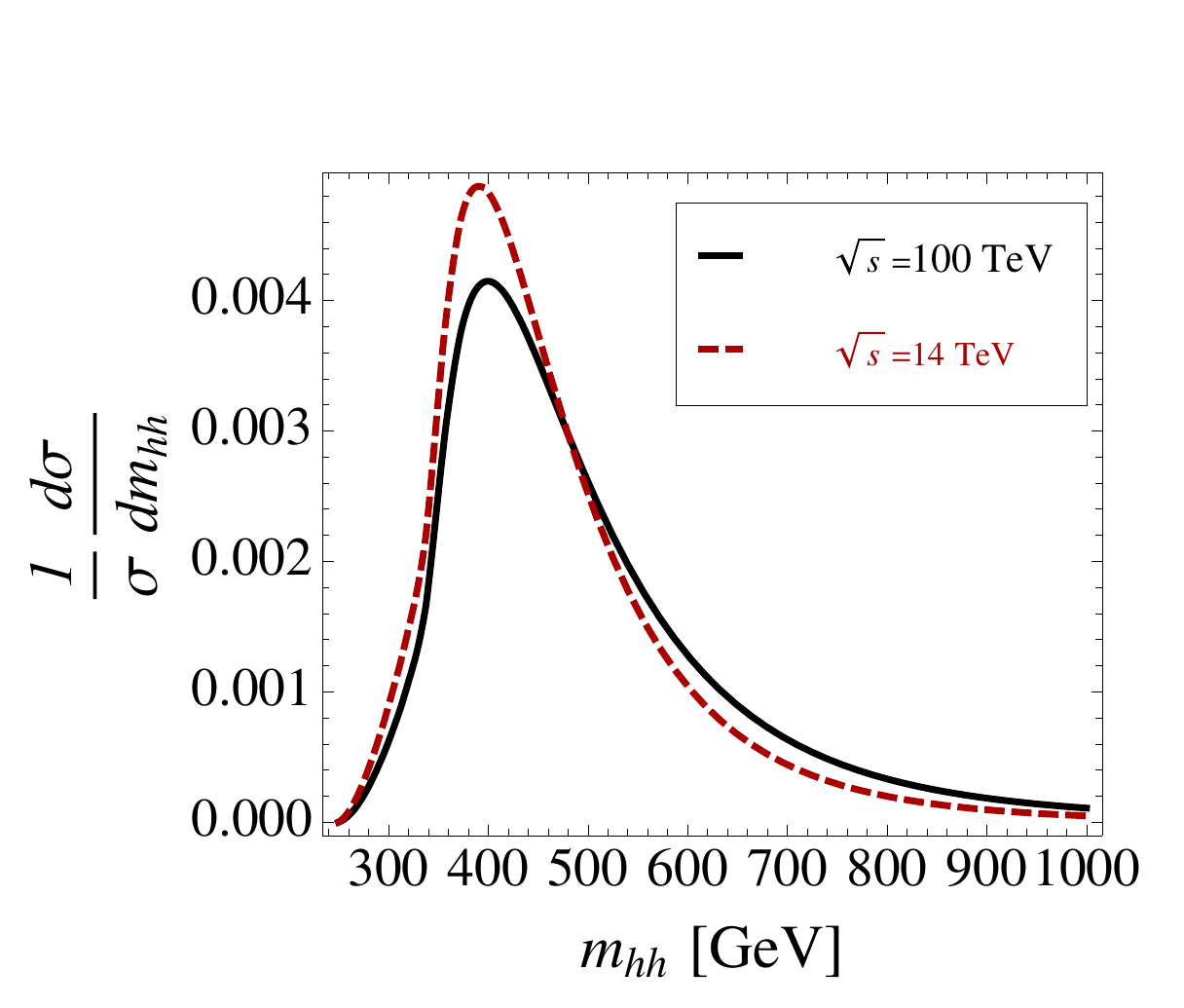}\label{fig2a}} 
\subfloat[]{\includegraphics[scale=0.68, angle=0]{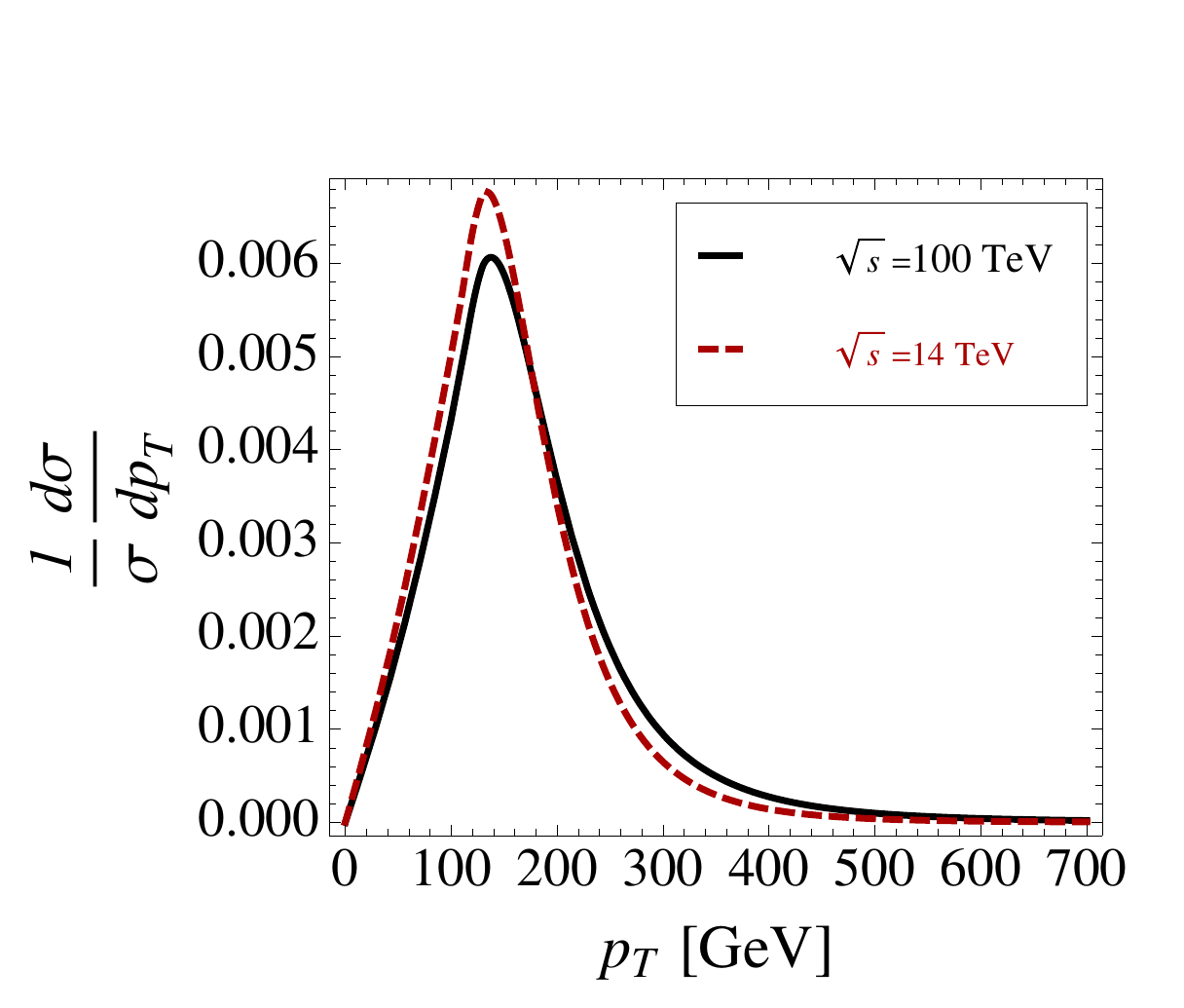}} 
\caption{\label{fig2}{\em Comparison of LO kinematic distributions in the SM at $\sqrt{s}=14$ and 100 TeV. }}
\end{figure}

 In Fig.~\ref{fig2} we show the LO $m_{hh}$ and $p_T$ distributions for SM $gg\to hh$ in a $pp$ collider at $\sqrt{s}=14$ and $100$ TeV. In this section we use {\tt LoopTools} \cite{Hahn:1998yk} to evaluate the loop functions in Eq.~(\ref{eq:amp}) and employ the MSTW 2008 LO 4F PDF \cite{Martin:2009iq}. Here all plots are produced this way with the following parameters:
\be
\label{eq:parameter}
m_t=173 {\rm \ \ GeV} \ , \quad m_h=125 {\rm \ \ GeV} \ , \quad \alpha_s^{\rm LO}(m_Z)=0.13355 \ .
\ee
We also set renormalization and factorization scales $\mu=m_{hh}$. It is clear that the overall shapes of these distributions are not sensitive to the CM energy of the hadron collider. The invariant mass distribution has a peak at $m_{hh} \sim 450$ GeV, while the $p_T$ distribution is maximum at $p_T \sim 150$ GeV.

\begin{figure}[t]
\subfloat[]{\includegraphics[scale=0.67, angle=0]{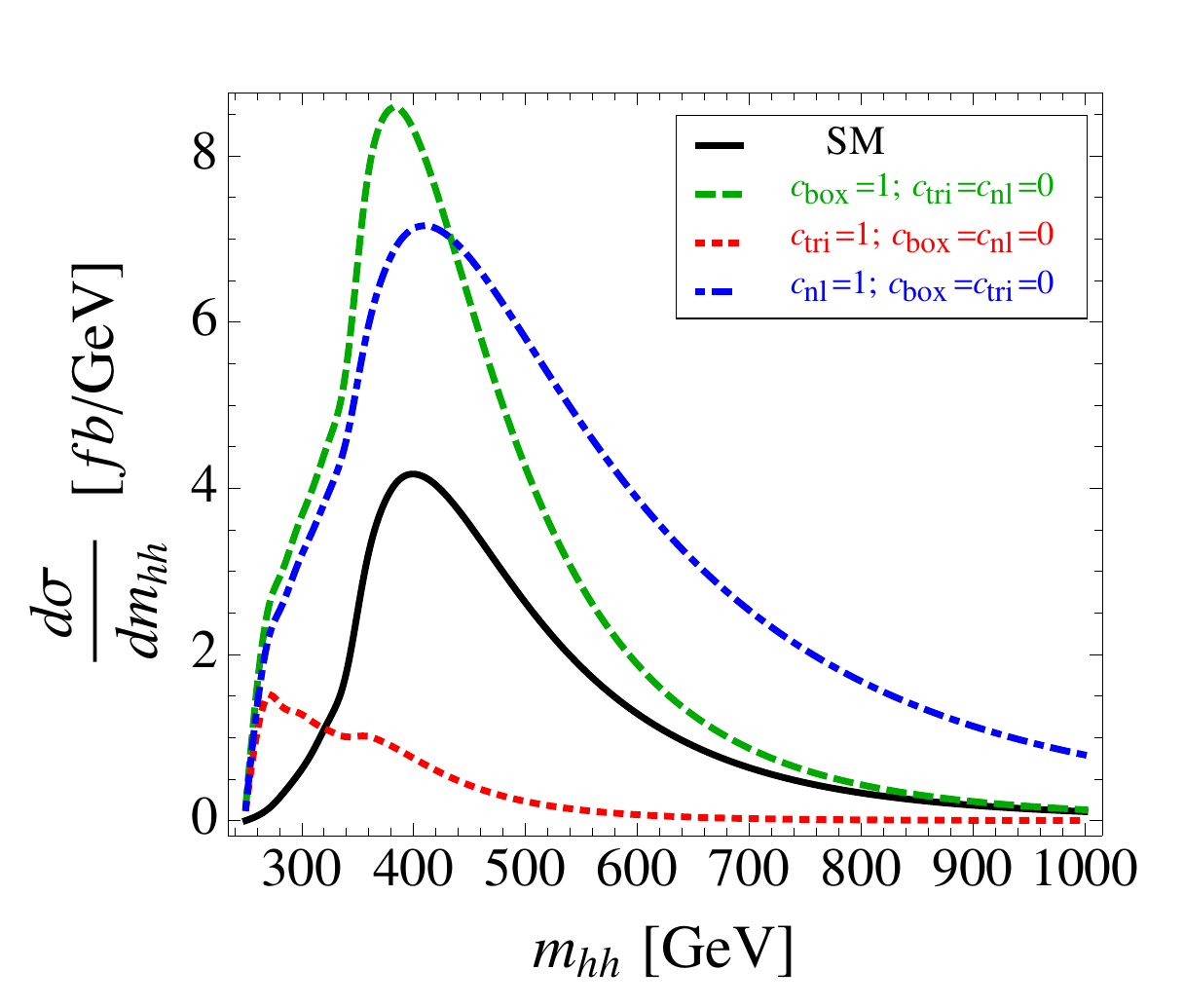}\label{fig3a}} 
\subfloat[]{\includegraphics[scale=0.69, angle=0]{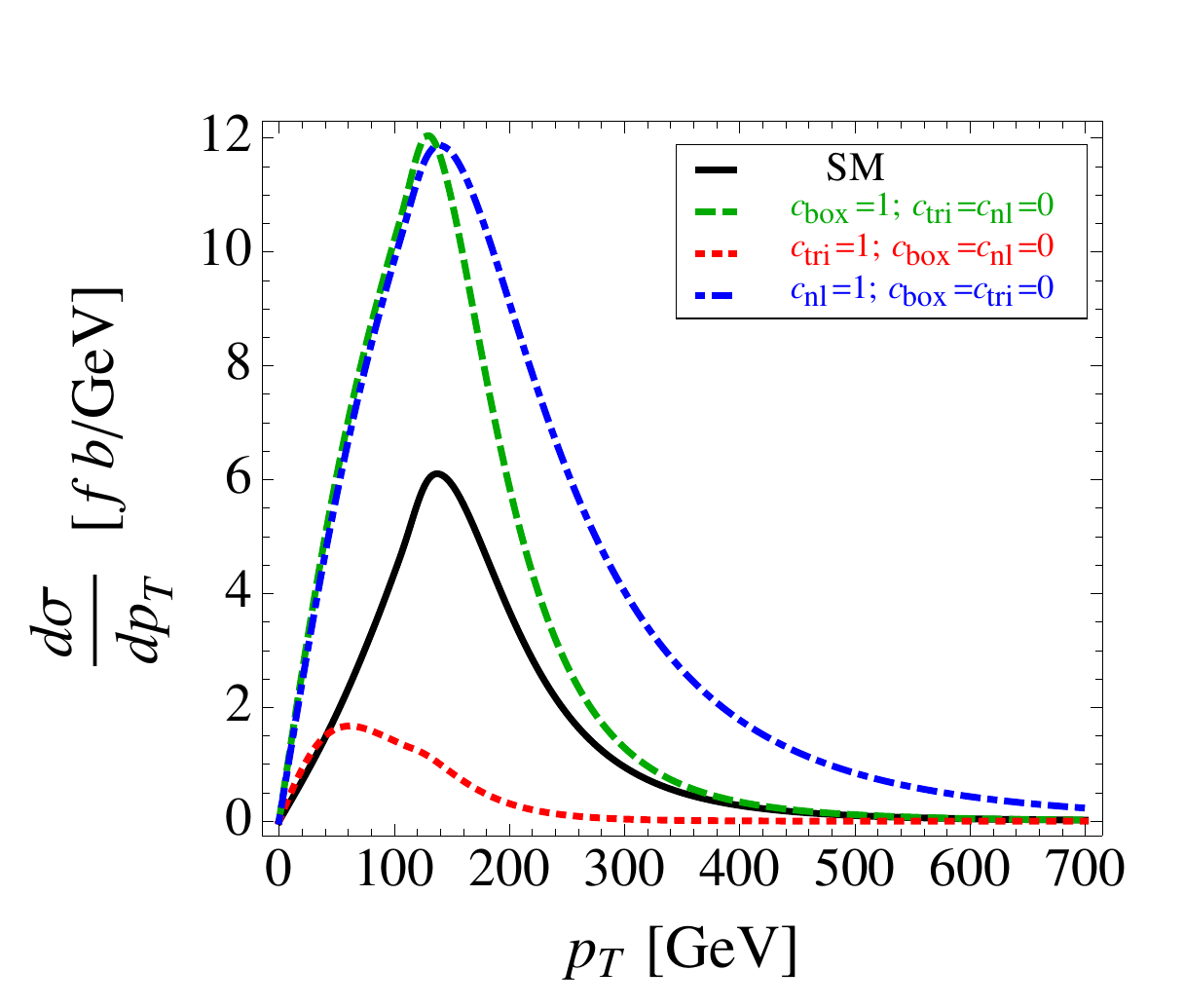}\label{fig3b}}
\caption{\label{fig3}{\em Individual contribution from $c_{tri}$, $c_{nl}$ and $c_{box}$, respectively, to the LO kinematic distributions in a $pp$ collider at $\sqrt{s}=100$ TeV.
}}
\end{figure}

From Fig.~\ref{fig2a} we see the majority of events have an invariant mass that is far above the kinematic threshold at $2m_h$. This observation has two important implications. The first is about the invalidity of the Higgs low-energy theorem in $gg\to hh$, which was already discussed in the end of Sect.~\ref{sect:new}. The second has to do with the relative weight between $c_{tri}$ and $c_{nl}$ in Eq.~(\ref{eq:ampara}), where the loop function $F_{\bigtriangleup}$ has the coefficient:
\be
c_{tri} \frac{3m_h^2}{\hat{s}-m_h^2} + c_{nl}\ .
\ee
Then we see that $c_{tri}$ becomes more important at small invariant mass, near the kinematic threshold $m_{hh} \sim 2m_h$, while $c_{nl}$ could easily dominate over $c_{tri}$ at large $m_{hh}$. In fact, since most of events have $m_{hh} \gg 2m_h$, the contribution from $c_{tri}$ will be suppressed in the total cross-section, which was the conclusion reached in Ref.~\cite{Contino:2012xk}. In other words, a truly model-independent measurement of the Higgs trilinear coupling from the total rate of $gg\to hh$ will be very difficult. In Fig.~\ref{fig3a} we show the individual contribution from $c_{tri}$, $c_{nl}$, and $c_{box}$, respectively, in the $m_{hh}$ distribution and compare them with the SM expectation. Indeed, when $c_{tri}=c_{nl}$ the contribution to the total cross section  from the Higgs trilinear coupling is quite small. As a result, turning on a small $c_{nl}$ would have a significant impact on the measurement of $c_{tri}$. From Fig.~\ref{fig3a} one could also infer that the interference between $F_{\bigtriangleup}$ and $F_{\Box}$ is destructive, a well-known observation.

Effects of new physics in the $p_T$ spectrum can be understood as follows. The loop functions $F_{\bigtriangleup}$ and $F_{\Box}$ represent contributions from initial gluons with the same helicity and have the angular momentum projection on the beam axis $J_z=0$, while $G_\Box$ arises from opposite helicity gluons and has $J_z=2$ along the beam axis \cite{Glover:1987nx, Plehn:1996wb}, which is why there is no interference between the two contributions in Eq.~(\ref{eq:amp}). Furthermore, $F_{\bigtriangleup}$ only contains $S$-wave orbital angular momentum since the Higgs couplings involved in Figs.~\ref{fig1}a and \ref{fig1}c are all scalar couplings and carry no angular momentum dependence. In other words, there is no $p_T$ dependence in $F_{\bigtriangleup}$ at all, which implies all the $p_T$ dependence in the $c_{tri}$ and $c_{nl}$ arise entirely from the phase space.  $F_\Box$, however, does carry the $J_z=0$ component of the $D$-wave angular momentum at higher order in the $\hat{s}/m_t^2$ expansion \cite{Dawson:2012mk}. Thus there is a residual $p_T$ dependence in $F_\Box$. Finally, $G_\Box$ has a strong $p_T$ dependence because of the $D$-wave nature.  In Fig.~\ref{fig3b} we show the $p_T$ spectrum from $c_{tri}$, $c_{nl}$ and $c_{box}$, turning on one parameter at a time. Similar to the $m_{hh}$ distribution, effects from $c_{tri}$ are suppressed in general, due to the off-shell propagator of the Higgs in Fig.~\ref{fig1}b.

\begin{figure}[t]
\subfloat[]{\includegraphics[scale=0.68, angle=0]{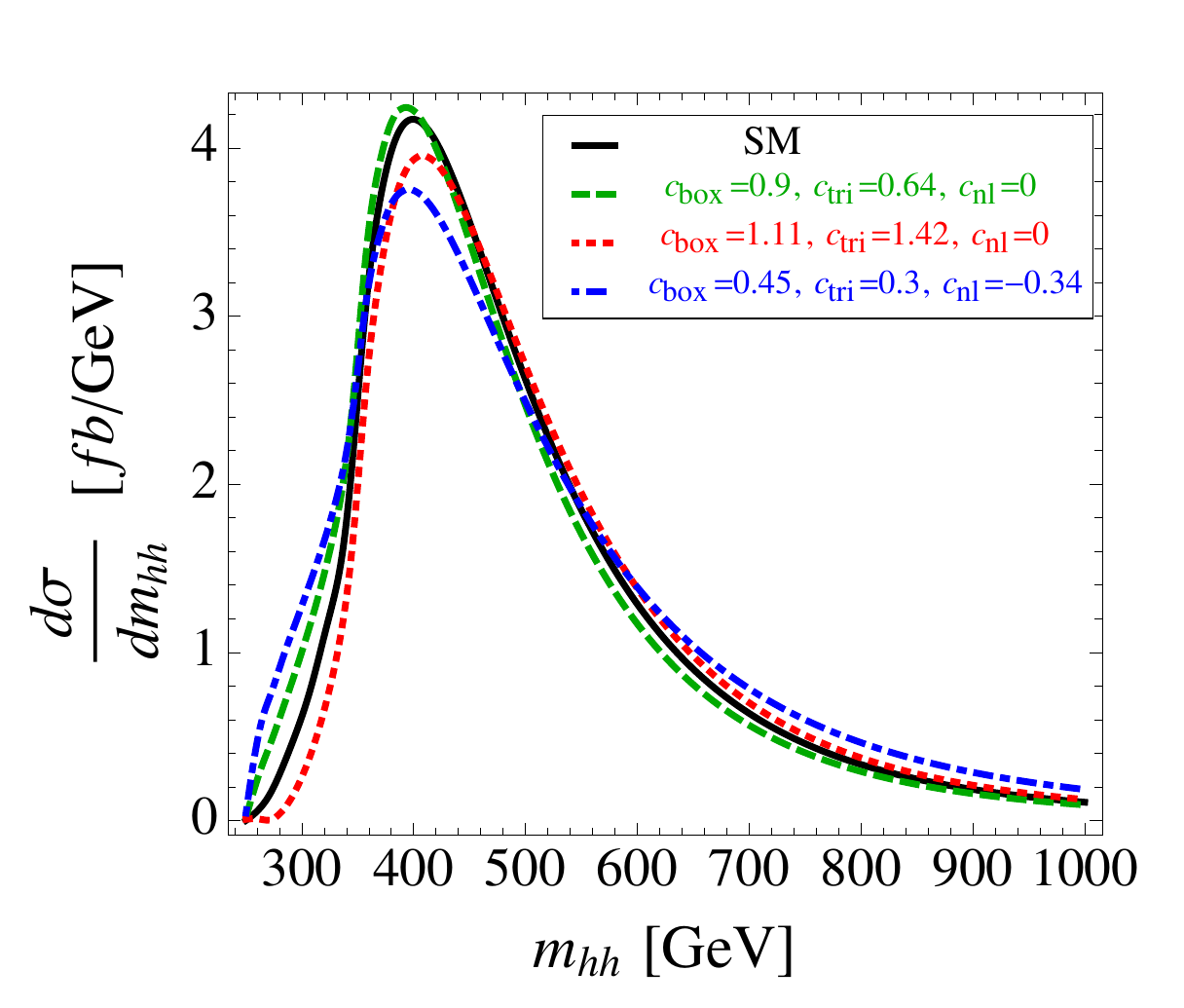}} 
\subfloat[]{\includegraphics[scale=0.69, angle=0]{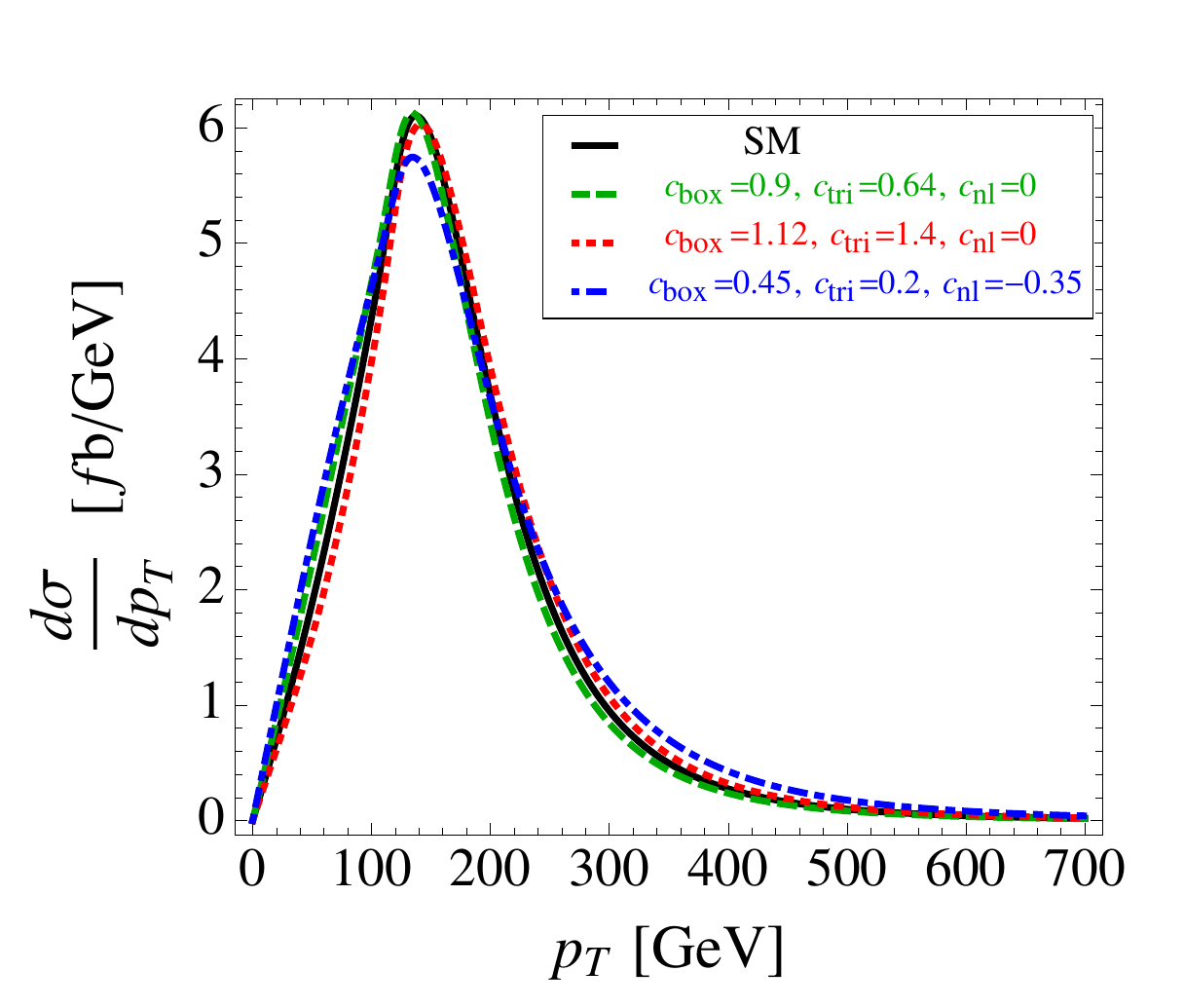}}
\caption{\label{fig4}{\em Similarities in kinematic distributions for various choices of $c_{box}$, $c_{tri}$, and $c_{nl}$ in a $pp$ collider at $\sqrt{s}=100$ TeV.
}}
\end{figure}

From Fig.~\ref{fig3} one can deduce a key result of the present study: even after including kinematic information in the $m_{hh}$ and $p_T$ distributions, various new physics contributions could still conspire to exhibit $m_{hh}$ and $p_T$ distributions that are similar to those expected in the SM. In Fig.~\ref{fig4} we show some choices of $c_{tri}$, $c_{nl}$ and $c_{box}$ which result in similar $m_{hh}$ and $p_T$ distributions. Fig.~\ref{fig4} also highlights the challenge of a precise measurement of the Higgs trilinear coupling using $gg\to hh$: a large number of events would be required to extract $c_{tri}$, $c_{nl}$ and $c_{box}$ and break the degeneracy among them. This is the motivation to base our Monte Carlo simulations and numerical analysis on future experiments in a 100 TeV $pp$ collider in the next Section.

\section{Simulations and Numerical Study}

In this section we perform numerical simulations of $gg\to hh$ in a 100 TeV $pp$ collider. We use the PYTHIA \cite{pythia} with the matrix elements from HPAIR~\cite{hpair, El-Kacimi:2002qba} and adopt CTEQ6L1 PDF~\cite{Pumplin:2002vw} to generate the events. 

First we consider effects of new physics in the total production rate of $gg\to hh$ before any event selections. In this case it is possible to parameterize the total rate in terms of the parameters $c_{tri}$, $c_{box}$ and $c_{tri}$, 
\bea
\label{eq:totalrate}
\sigma(gg\to hh) = \sigma^{SM}(gg\to hh) [ 1.849~ c_{box}^2 + 0.201~ c_{tri}^2 + 2.684~ c_{nl}^2  \nonumber  \\
- 1.050~ c_{box} c_{tri} - 3.974~ c_{box}  c_{nl} +1.215~ c_{tri} c_{nl} ].
\eea 
By comparing with a similar result in Ref.~\cite{Contino:2012xk} for the LHC with $\sqrt{s}=14$ TeV, we see at $\sqrt{s}=100$ TeV there is not much change in the numerical coefficients in the above equation. In particular, the coefficient of $c_{tri}^2$ is an order of magnitude smaller than those of $c_{box}^2$ and $c_{nl}^2$, a crucial observation already made in Ref.~\cite{Contino:2012xk}. 
%
\begin{figure}[t]
\subfloat[]{\includegraphics[scale=0.3, angle=0]{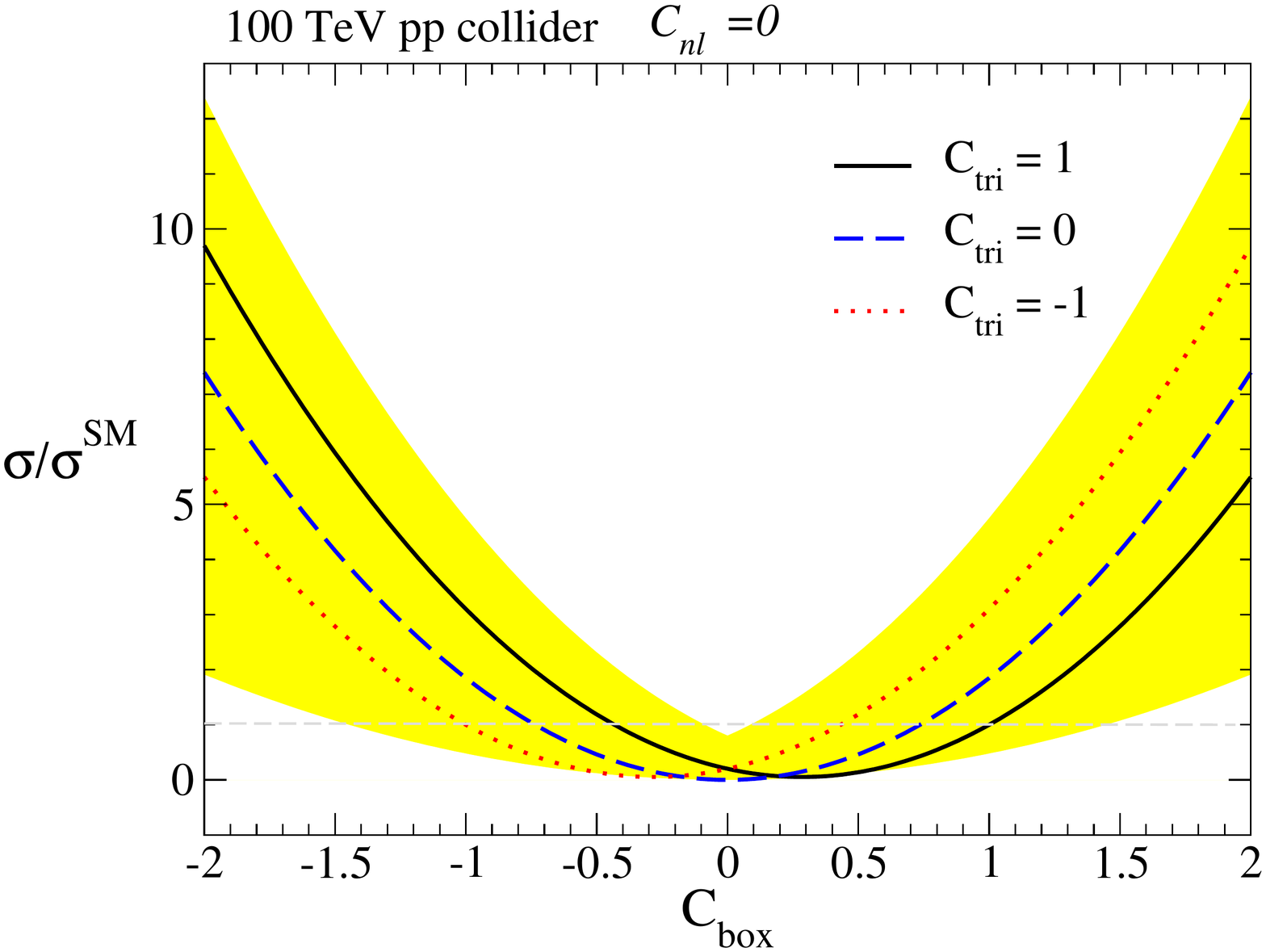}\label{fig:xs_11}} 
\subfloat[]{\includegraphics[scale=0.3, angle=0]{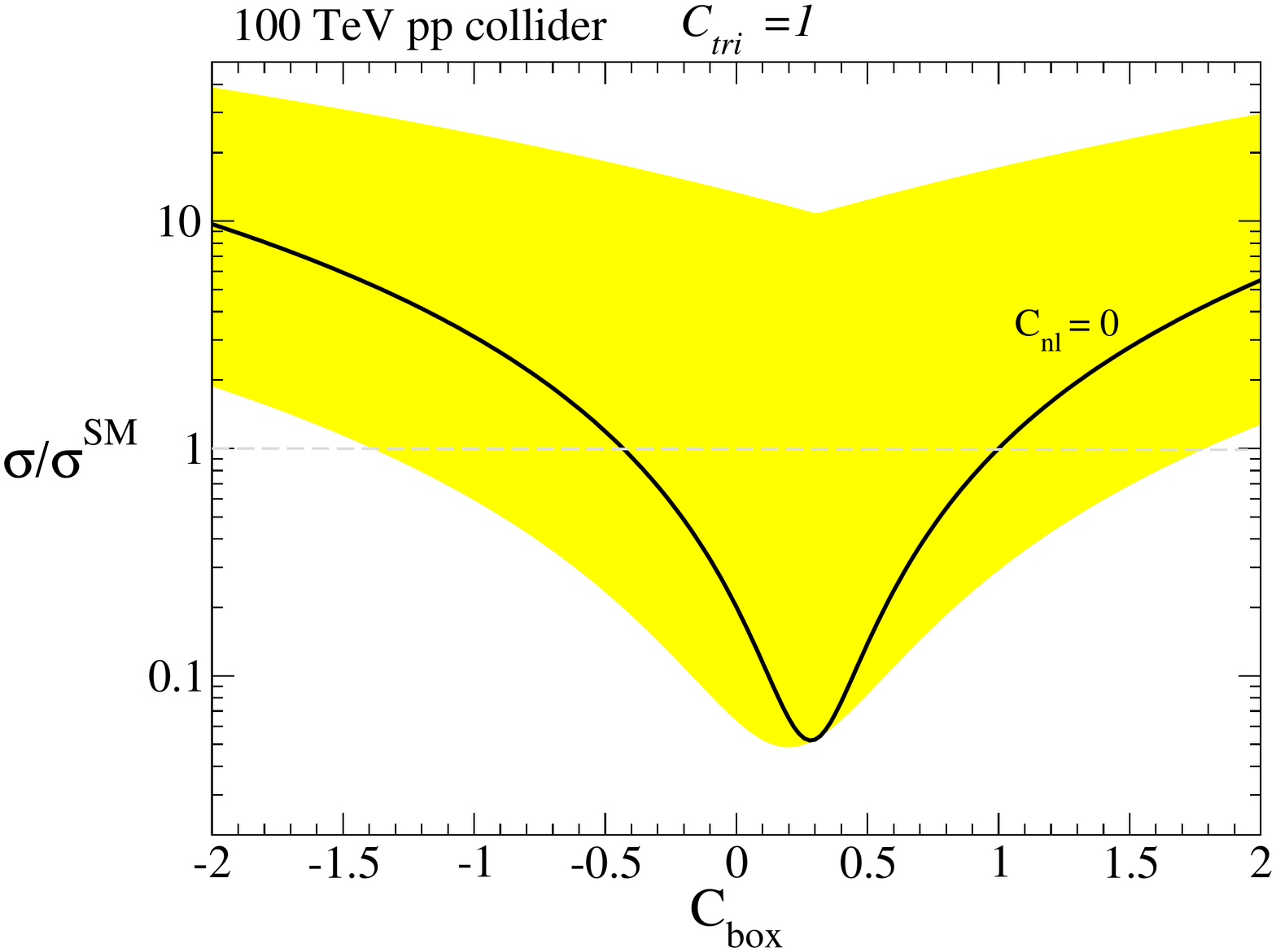}\label{fig:xs_12}}
\caption{\label{fig:xs_1}{\em {\rm (a)} The yellow region shows $\sigma/\sigma^{SM}$ by varying $c_{box}$ and $c_{tri}$ between $-2$ to $2$ and setting $c_{nl}=0$. 
The horizontal line indicates no deviation from the SM rate.  {\rm (b)} Same as {\rm (a)}, but with $c_{tri}=1$ and $c_{nl}$ varying from $-2$ to $2$. 
}}
\end{figure}

Employing Eq.~(\ref{eq:totalrate}), we show in Fig.~\ref{fig:xs_1} some examples of new physics effects in the ratio of  the total production cross section of $gg\to hh$ over the SM expectation. In Fig.~\ref{fig:xs_11} $c_{nl}$ is turned off while $c_{box}$ and $c_{tri}$ are both allowed to vary between $-2$ and 2. The resulting variation in the total rate is shown in the yellow band, which shows strong enhancement when $|c_{box}|\gtrsim 1.5$, and the enhancement can be as large as a factor of $10$ when $c_{box}=\pm2$. In the plot we also show three reference cases for $c_{tri}=1$, $c_{tri}= 0$ and $c_{tri} = -1$ with black-solid, blue-dashed and red-dotted curves, respectively.  
It is clear that a significant region of  the parameter space in $c_{box}$ and $c_{tri}$ could conspire to produce the same cross section of $gg\to hh$ as in the SM, even though the trilinear coupling of Higgs boson vanishes or has an opposite sign to the SM.  In Fig.~\ref{fig:xs_12}, we fix $c_{tri}$ to be unity, its SM value, and study the effects caused by varying $c_{box}$ and $c_{nl}$ between $-2$ and $2$. The production cross section is always enhanced when $c_{box} \gtrsim 1.8$ or $c_{box}\lesssim -1.4$ and can be a factor of $40$ larger than the SM when $c_{box}=-2$. The black curve in Fig.~\ref{fig:xs_12} is for a vanishing $c_{nl}$ that reproduces the corresponding black-solid curve in Fig.~\ref{fig:xs_11}. Again, a significant region of the parameter space in $c_{box}$ and $c_{nl}$ could give rise to the SM total rate in $gg\to hh$.

\begin{figure}[!t]
\subfloat[]{\includegraphics[scale=0.26, angle=0]{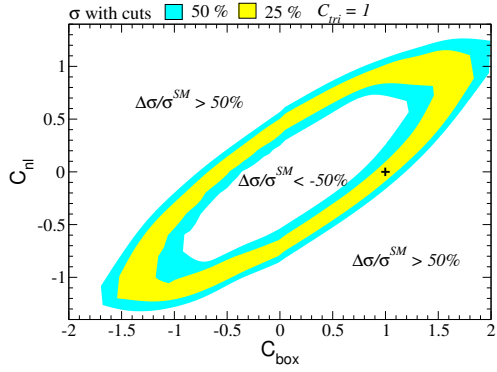}\label{fig:xs_cut1}} \\ \vspace{-0.6cm}
\subfloat[]{\includegraphics[scale=0.26, angle=0]{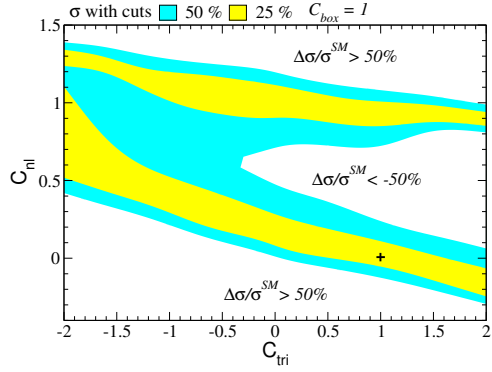}\label{fig:xs_cut2} }\\ \vspace{-0.6cm}
\subfloat[]{\includegraphics[scale=0.26, angle=0]{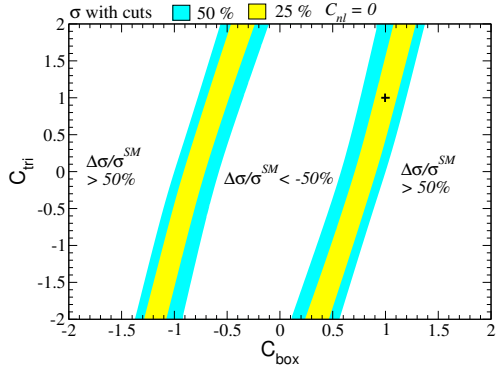}\label{fig:xs_cut3}}
\caption{\label{fig:xs_cut}{\em Contour plot for the cross section of $gg\to hh\to b\bar{b}\gamma\gamma$ after imposing Eq.~(\ref{eq:cuts}). The yellow and cyan bands indicate the parameter space that agree with the SM result within $25\%$ and $50\%$, respectively. The SM value is marked with a black cross.
}}
\end{figure}

Next we study the impact of event selections on extracting new physics effects in the double Higgs production. In a $100$ TeV $pp$ collider, it was shown that~\cite{Baglio:2012np, Yao:2013ika}  the process $gg\to hh$ can be discovered in $b\bar{b}\gamma\gamma$ channel. Following Refs.~\cite{Baglio:2012np, Yao:2013ika} we impose the following event selections:
\bea
p_T^b > 35~{\rm GeV},~~|\eta_b|<2,~~2.5>\Delta R(b,b) > 0.4,\nonumber \\
p_T^\gamma >35~{\rm GeV},~~|\eta_\gamma|<2,~~2.5>\Delta R(\gamma,\gamma)> 0.4,~~\Delta R(\gamma,b) > 0.4,\nonumber \\
|\cos\theta_{\gamma\gamma}|<0.8, ~~p_T^h > 100~{\rm GeV~and}~ m_{hh} > 350~{\rm GeV},\nonumber
\label{eq:cuts}
\eea 
where $\theta_{\gamma\gamma}$ is the angle between two photons in the rest frame of two Higgs bosons. In this case we find a simple parameterization like Eq.~(\ref{eq:totalrate}) cannot apply anymore, for the selection efficiency would depend on the parameters $c_{box}$, $c_{tri}$ and $c_{nl}$, which should be obvious from the fact that the kinematic distributions also depend on these parameters.

In Fig.~\ref{fig:xs_cut} we consider  constraints on the $c_{box}$, $c_{tri}$ and $c_{nl}$ from the total rate measurements at $\sqrt{s}=100$ TeV, by assuming 25\% and 50\% deviations from the SM expectation, respectively. In each plot in Fig.~\ref{fig:xs_cut}, we fix  one of $c_{tri},~c_{box}$ and $c_{nl}$ to be the value in the SM and vary the other two. The yellow band indicates the parameter space that agrees with the SM result within $25\%$, while the cyan band represents the region for $50\%$.  More specifically, in Fig.~\ref{fig:xs_cut1}, where $c_{tri}=1$ takes the SM value, both $c_{box}$ and $c_{nl}$ can be constrained within the interval $(-2,2)$, roughly speaking. Moreover, because the triangle diagram interferes destructively with the box diagram, any effect from increasing $c_{box}$ can be offset by increasing $c_{nl}$ as well.
Next assuming a SM $c_{box}=1$ in Fig.~\ref{fig:xs_cut2}, we see explicitly the total rate has poor sensitivity to $c_{tri}$, which involves the Higgs trilinear coupling. This insensitivity persists in Fig.~\ref{fig:xs_cut3}, where we set $c_{nl}=0$ as in the SM. These findings strongly motivate searching for additional kinematic information to unravel the various new physics contributions in the double Higgs production, which we turn to in the following.

\begin{figure}[!t]
\subfloat[]{\includegraphics[scale=0.26, angle=0]{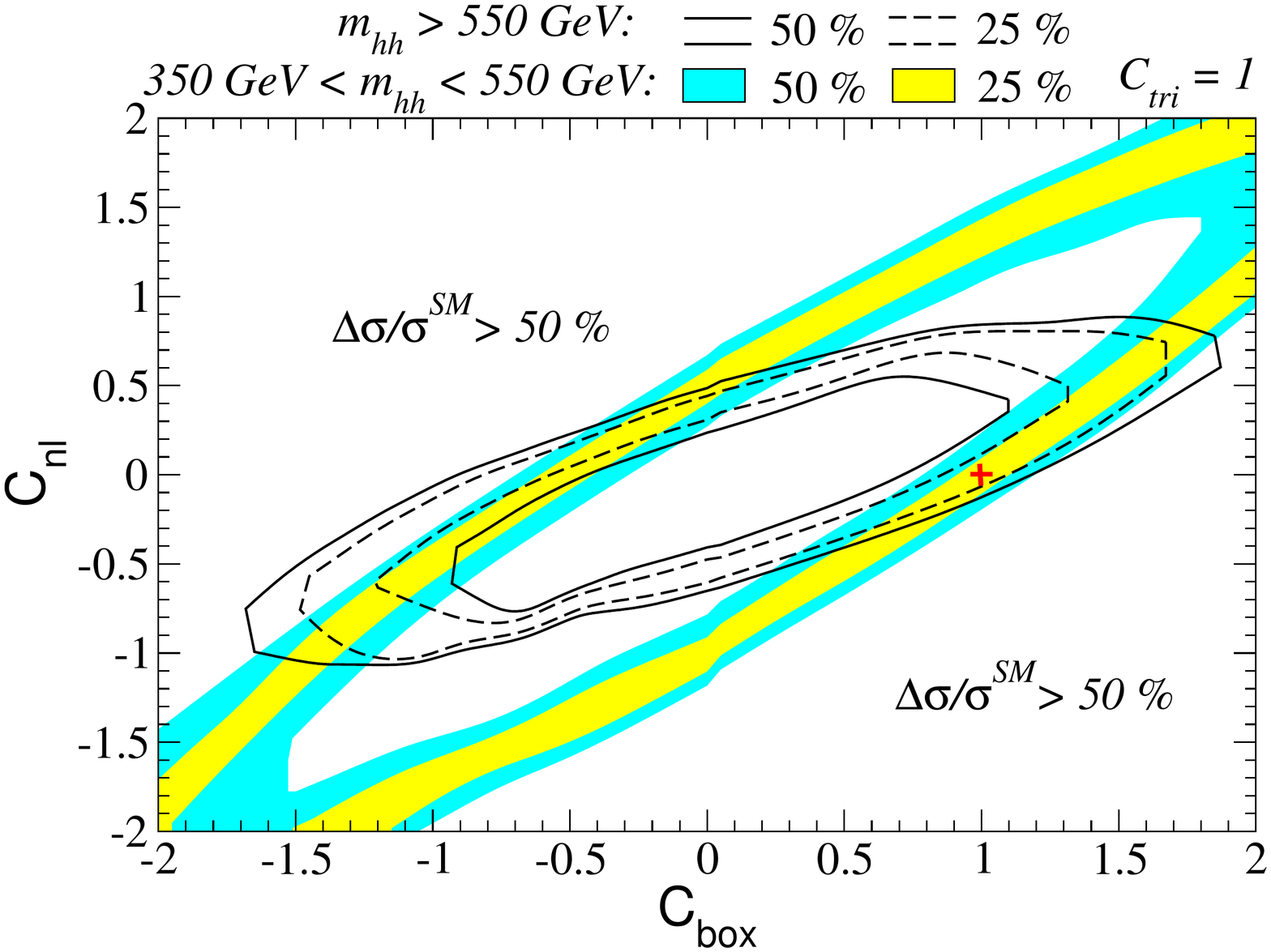}\label{fig:mhhcut1}} \\ \vspace{-0.6cm}
\subfloat[]{\includegraphics[scale=0.26, angle=0]{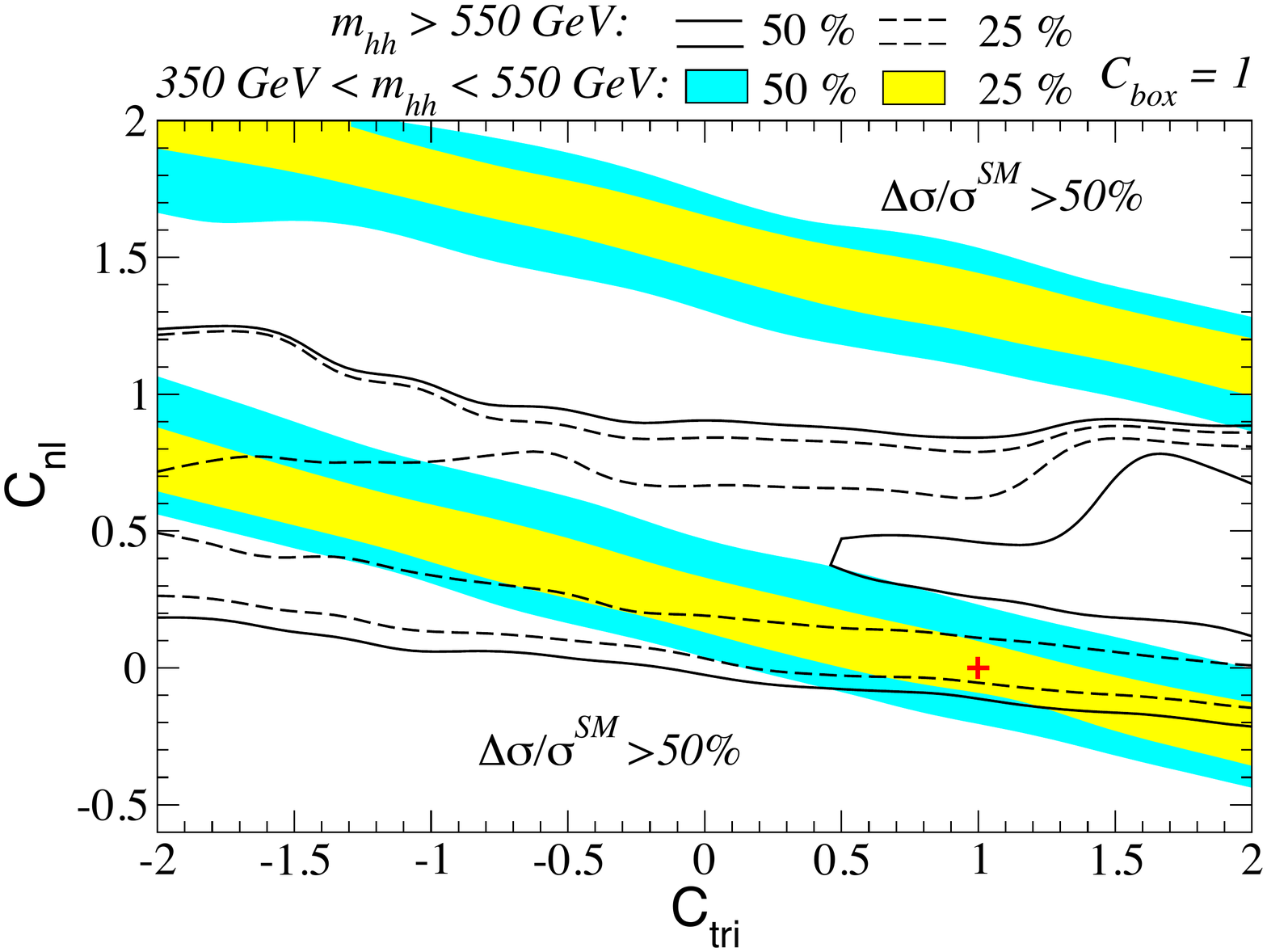}\label{fig:mhhcut2}} \\ \vspace{-0.6cm}
\subfloat[]{\includegraphics[scale=0.26, angle=0]{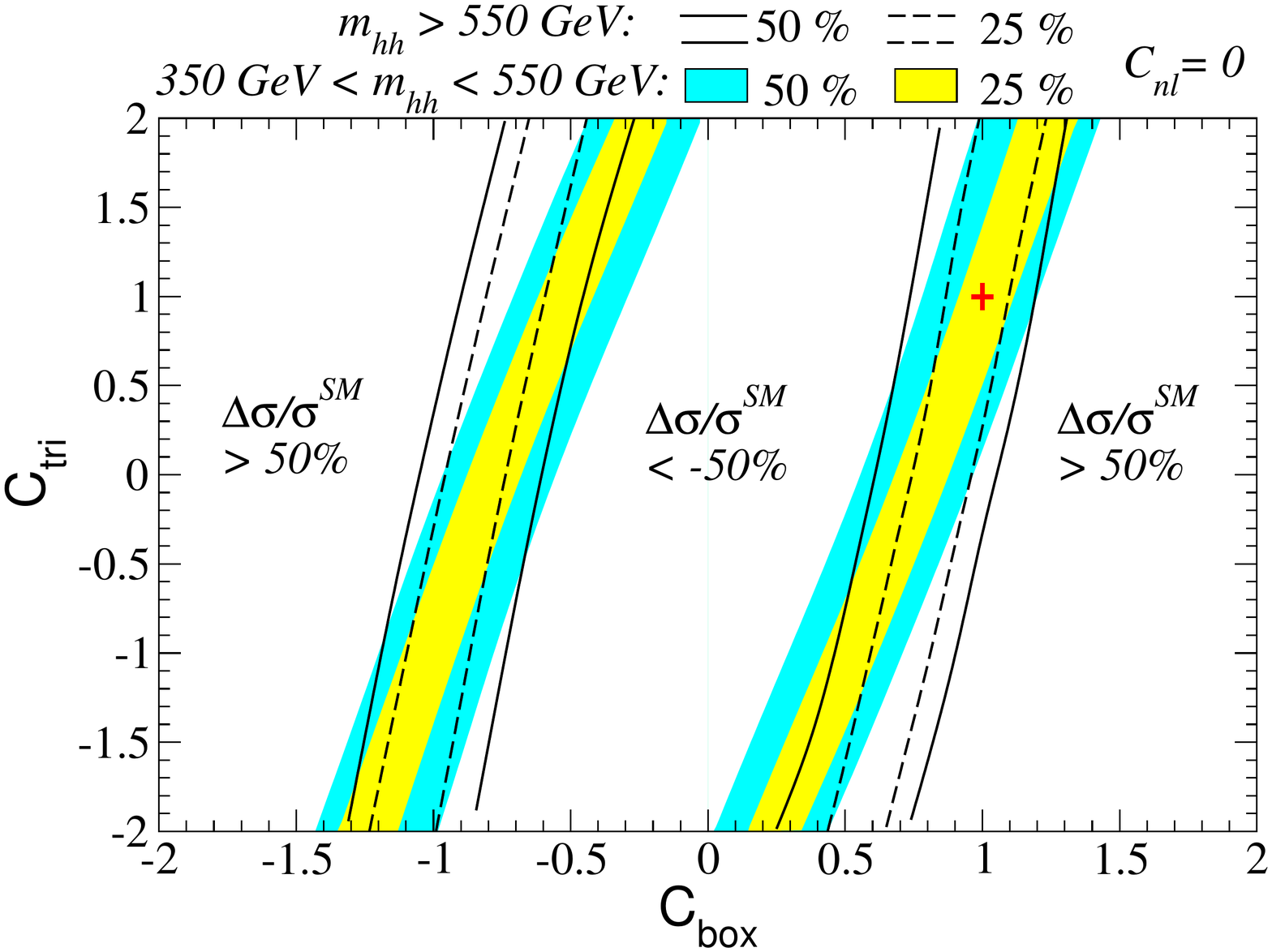}\label{fig:mhhcut3}}
\caption{\label{fig:mhhcut}{\em Contour plots for the cross section in two energy bins. Bin I: $350~{\rm GeV} < m_{hh} < 550~{\rm GeV}$ and Bin II: $m_{hh}>550~{\rm GeV}$. The yellow (cyan) band and the region with two dashed (solid) black curves are consistent with SM results within $25\%$ ($50\%$) for Bin I and Bin II, respectively. The SM value is marked with a red cross.
}}
\end{figure}

As we have seen in the previous section, contributions from $c_{box}$, $c_{tri}$ and $c_{nl}$ have somewhat different distributions in transverse momentum $p_T^h$ and invariant mass of two Higgs bosons $m_{hh}$: the $c_{tri}$ component peaks at low $m_{hh}$, the $c_{box}$ piece shifts $m_{hh}$ to higher vales, and the $c_{nl}$ coupling pushes the distribution to even larger $m_{hh}$. (See Fig.~\ref{fig3}). As a first step toward including the kinematic information in the differential spectra, we divide the $m_{hh}$ and $p_T$ distributions into two bins: a low bin and a high bin. The differential rate in each bin is then used to constrain $c_{box}$, $c_{tri}$ and $c_{nl}$. In so doing we find the constraints from fitting the two $p_T$ bins are quite similar to those from fitting the two $m_{hh}$ bins. Therefore, in what follows we only present the constraints from fitting the low and the high $m_{hh}$-bins. 

From Fig.~\ref{fig3a} we choose the following two $m_{hh}$ bins in our analysis:
\bea
{\rm Bin\ \ I}&:& 350~{\rm GeV} \le m_{hh} \le 550~{\rm GeV} \nonumber \\
{\rm Bin\ II}&:& 550 ~{\rm GeV}  \le m_{hh} \nonumber
\eea
We then consider the constraints by allowing the differential rate in each bin to fall within 25\% and 50\% of SM expectations, which are shown in Fig.~\ref{fig:mhhcut}. Again in each plot in Fig.~\ref{fig:mhhcut} one  of the $c_{box}$, $c_{tri}$ and $c_{nl}$ is chosen to be the SM value while the other two are allowed to vary. In Fig.~\ref{fig:mhhcut1}, where $c_{tri}=1$, we see measurements in the two bins could break the degeneracy in $c_{box}$ and $c_{nl}$ effectively, as the two sets of contours from Bin I and Bin II have only a small region of overlap. However we caution that some degeneracy still remains even if the differential rates in the two bins both conform to SM expectations. The situation becomes worse when it comes to constraining $c_{tri}$. In Figs.~\ref{fig:mhhcut2} and \ref{fig:mhhcut3} where $c_{tri}$ is allowed to vary, along with one other parameter, we see the overlap from two sets of contours become larger than in Fig.~\ref{fig:mhhcut1}. Nevertheless, the inclusion of kinematic information from these two $m_{hh}$ bins still allow for a significant improvement in constraining $c_{tri}$ from using the total rate measurement alone.

\section{conclusions}

In this work we initiated a study on using the kinematic distribution  to disentangle new physics effects in $gg\to hh$, which is the dominant channel to extract the Higgs trilinear coupling. Parameterizing the different new physics effects in the differential cross section in terms of three dimensionless coefficients, $c_{box}$, $c_{tri}$ and $c_{nl}$, we studied the interplay of these different contributions in the $p_T$ and total invariant mass spectra of the Higgs bosons. Next we performed a numerical study of constraining these parameters in a future 100 TeV $pp$ collider by fitting the differential rates in a low invariant mass and a high invariant mass bins. Constraints from a low $p_T$ and a $p_T$ bins turned out to be very similar to those from the two invariant mass bins. In the end, we found $c_{box}$ and $c_{nl}$ could be constrained effectively, although some degeneracy survives. On the other hand, the constraint on $c_{tri}$, which includes the effect of the Higgs trilinear coupling, remains quite weak. Nevertheless, using the kinematic information from the two invariant mass bins still shows significant improvements from using the total production rate alone.

Given that self-interactions of the Higgs boson is the only aspect of the 125 GeV Higgs boson that has not been tested experimentally, measurements on the Higgs trilinear coupling should be among the highest priorities in future research programs on properties of the Higgs boson. Our work is only  a first step toward precision measurements on the Higgs self-interactions. To be able to make use of the full kinematic information, ideally one would like to perform a multivariate analysis based on the Matrix Element Method \cite{Kondo:1988yd}, which has been applied to the top quark analyses \cite{Dalitz:1991wa} and the Higgs discovery in the $4\ell$ channel \cite{Gao:2010qx}. We plan to continue to pursue this direction in a future study.

\begin{acknowledgments}
The work of C.-R.C.  is supported in part by the National Science Council of R.O.C. under Grants No.~NSC 102-2112-M-003-001-MY3 . 
I.L.~is supported in part by the U.S. Department of Energy under Contracts No. DE-AC02- 06CH11357 and No. DE-SC0010143. I.L. would like to acknowledge the hospitality at Centro de Ciencias de Benasque Pedro Pascual, where part of this work was performed.
\end{acknowledgments}


\end{document}